\documentclass[aps,prl,twocolumn,groupedaddress,showpacs,amsmath,final]{revtex4}
\usepackage{graphicx}
\begin{document}

\title{Measurement of the shot noise in a single electron transistor.}

\author{S. Kafanov\footnote{Present address: Low Temperature Laboratory, Helsinki University of Technology, P.O. Box 3500, 02015 TKK, Finland.}}
\email{sergey.kafanov@ltl.tkk.fi}
\author{P. Delsing}
\email{per.delsing@chalmers.se}
\affiliation{Microtechnology and Nanoscience, Chalmers University of
Technology; S-41296, G\"{o}teborg, Sweden.}
\date{\today}

\begin{abstract}
We have systematically measured the shot noise in a single electron transistor (SET) as a function of bias and gate voltages. By embedding a SET in a resonance circuit we have been able to measure its shot noise at the resonance frequency $464\,\mathrm{MHz}$, where the $1/f$ noise is negligible. We can extract the Fano factor which varies between $0.5$ and $1$ depending on the amount of Coulomb blockade in the SET, in very good agreement with the theory.
\end{abstract}


\maketitle

In both electronic and photonic devices the measured signal does not contain all the information about the state and dynamics of a system. In most cases there is additional information in the fluctuations of the signal \cite{Nature.392.358}.
For example  the shot noise in electronic circuits contains information about the charge of charge carriers. This was used by Saminadayar \textit{et al.} to demonstrate the fractional charge of the quasiparticles in a fractional quantum Hall system \cite{PhysRevLett.79.2526}. The shot noise can also reveal correlations of the charge carriers or photons. Bosons display bunching  \cite{Nature.178.1447} whereas fermions display anti-bunching due to the Pauli principle \cite{Science.284.296, Science.284.299}. 

Electron tunneling in a tunnel junction or a point contact is described by a Poissonian process. 
If the transparency of the barrier $T$ is small,  the shot noise is proportional to $T$
and the shot noise takes on the Schottky value $S=2e\langle I \rangle$, where $\langle I \rangle$ is the average current. 
To compare different situations the noise is often normalized to the Schottky value giving the so called Fano factor $F=S/(2e\langle I \rangle )$ \cite{PhysRev.72.26}. The Fano factor can be reduced for example in a quantum point contact where the trasparency is large, then the shot noise is  proportional to $T(1-T)$ for each of the participating channels \cite{Kumar, Reznikov}.
Another examle is an array of identical series connected tunnel junctions, there the shot noise is inversely proportional to the numbers of junctions $N, S=2e \langle I \rangle /N$ \cite{PhysRevB.50.17674}. The Fano factor can also be enhanced above unity which is the case when cotunneling dominates.

In a Single Electron Transistor (SET) \cite{Magn.IEEE.Trans.23.1142, PhysRevLett.59.109}, where $N=2$, the tunneling is uncorrelated at high bias, and the shot noise is $S=e\langle I \rangle$, \textit{i.e.} the Fano factor is $1/2$. However, at low voltage the Coulomb blockade introduces correlations and the Fano factor increases. The correlation of tunnel events in a SET depend strongly on the biasing conditions and the amount of Coulomb blockade in the SET.

The shot noise in SETs is theoretically well understood \cite{Korotkov, PhysRevB.49.10381, PhysRevB.47.1967, Hanke},
however, so far a detailed comparison between experiment and theory has not been demonstrated. This is due to the fact that the shot noise in most cases is masked by other sources of noise. At low frequency, motion of background charge fluctuators give rise to $1/f$ noise \cite{ApplPhysLett.61.237}, at higher frequencies the amplifier noise normally dominates over the shot noise. In a few cases the shot noise of double junction devices without a gate has been measured \cite{PhysRevLett.75.1610, PhysRevB.66.161303, PhysRevB.70.033305}. In other experiments on silicon and carbon nanotube SETs the shot noise was studied with gate control. Sasaki \textit{et al.} \cite{Sasaki} studied a silicon SET and found a Fano factor lower than one half in the Coulomb blockade region in contradiction with theory. Onac \textit{et al.} \cite{Onac} studied a carbon nanotube SET and found super-poissonian shot noise with a Fano factor larger than unity, however there was no quantitative comparision to theory.
Metallic SETs have been studied at high bias \cite{PhysRevLett.86.3376, ApplPhysLett.79.4031, JApplPhys.95.1274}, but the low bias Coulomb blockade region where the correlations affect the Fano factor has so far not been studied. Metallic SETs in the superconducting state has recently been studied \cite{Xue}.

In this paper we present systematic measurements of the shot noise of a SET for different gate and bias voltages. We make a quantitative comparison with theory and show how the Fano factor varies between 0.5 and 1 depending on the correlations in the SET, in very good agreement with theory.
We do this using a rf-SET set-up \cite {Science280.1238} where a current is fed through the SET and the noise is coupled out to a cold amplifier via a resonance circuit.
In principle the shot noise is frequency dependent
\cite{PhysRevB.49.10381}, however since our measurements are done at a frequency much lower than the characteristic SET frequency
($1/(2\pi R_iC_i) \sim 20\,\mathrm{GHz}$ we are still in the low frequency limit. Here $R_i$ and $C_i, i \in \{1,2\}$ are the resistances and capacitances of the two junctions.
\begin{figure}[t]
\includegraphics[width=0.95\columnwidth]{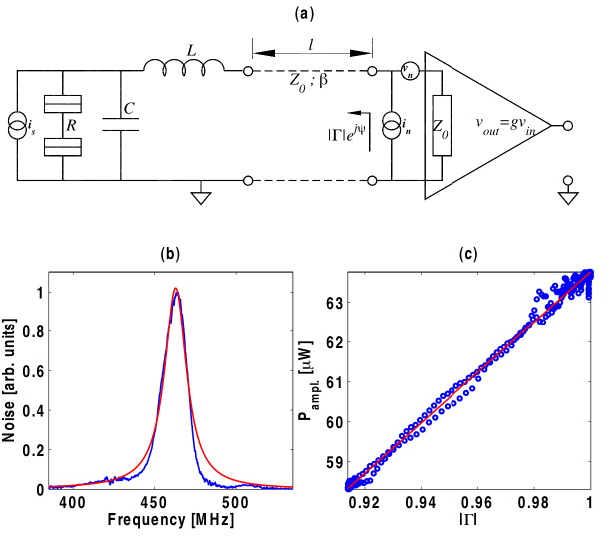}
\caption{{\label{figure1}}\small (color online)(a) Simplified
circuit scheme for the shot noise measurements with a rf-SET. The
resonance circuit is connected to the microwave amplifier via a transmission line with a characteristic impedance
$Z_0$ and propagation constant $\beta$.
(b) Resonance curve for the resonance circuit. The blue curve is the measured noise power as a function of frequency. The red curve is a fit to eq.\,\ref{Resonance_Curve}. (c) The measured amplifier noise as a function of the magnitude of the reflection coefficient. The red curve is a fit to eq.\,\ref{AmplNoise}}
\end{figure}

The sample was fabricated from aluminum on an oxidized silicon substrate using electron beam lithography and a standard double-angle evaporation technique. The sample was attached to the mixing chamber of a dilution refrigerator and cooled to a temperature of $\sim 25\,\mathrm{mK}$. All measurements were performed in the normal (nonsuperconducting) state at a magnetic field of $1.5\,\mathrm{T}$. The SET was biased in the asymmetric mode with one lead  grounded. The high bias asymptotic resistance of the SET was $R\simeq 72\,\mathrm{k\Omega}$. The charging energy,
$E_C/k_B=e^2/(2k_\mathrm{B}C_\Sigma)\simeq 2.5\,\mathrm{K}$ was extracted from the measurements of the SET stability diagram, where $C_\Sigma$ is the sum capacitance of the SET island. The asymmetry in the junction capacitances $\sim 30\%$ was deduced from the asymmetry of the SET stability diagram.

We have embedded the SET in an LC-resonance circuit, which transforms the high SET resistance to an impedance close to $50\,\mathrm{\Omega}$.
The resonance circuit is connected via a superconducting coax to a low noise cryogenic amplifier situated at $4.2\,\mathrm{K}$. The coax is characterized by a wave impedance $Z_0=50\,\Omega$ and a propagation constant $\beta$. We have assumed that the input and output impedances of the cryogenic amplifier are equal to $Z_0$. The output signal from the cold amplifier is further amplified at room temperature, and recorded with a spectrum analyzer. 

The parameters of the matching circuit can be obtained from reflectometry measurements. A small rf-signal at the resonance frequency $f_{LC}$ is launched toward the SET and part of this signal is reflected from the circuit. The reflection coefficient, $\Gamma=(Z-Z_0)/(Z+Z_0)$, is the ratio between the voltages of the reflected and the incoming wave, and depends on the impedance mismatch between the resonance circuit impedance $Z$ and $Z_0$. The resonance frequency of the resonance circuit is defined by the LC parameters, $f_{LC} = 1/(2 \pi \sqrt{LC}$) and was measured to $464\,\mathrm{MHz}$ (see Fig.\ref{figure1}b).


The shot noise from the SET can be described as a current noise source $i_s$ in parallel with the SET, as shown in Fig.\,\ref{figure1}a. The power spectral density of the shot noise is given by the Fano factor $F$ and the current through the SET, $\langle i_s^2 \rangle=2e \langle I \rangle F$. 
Considering the electrical scheme in Fig.\,\ref{figure1}a, and taking into account that $R\gg Z_0$, we can calculate the power spectral density  at the amplifier output, resulting from the shot noise of the SET, as a function of the resonance circuit parameters:
\begin{equation}{\label{Resonance_Curve}}
S_{shot}=\frac{g^2Z_0\langle i_s^2\rangle}{(1-(f/f_{LC})^2)^2+(f/f_{LC})^2(1/Q)^2},
\end{equation}
where $g$ is the amplifier voltage gain and $Q=(Z_{LC}/R+Z_0/Z_{LC})^{-1}$ is the total quality factor of the resonance circuit.
$Z_{LC}=\sqrt{L/C}$ is the characteristic impedance of the LC-circuit (not including R).
The detected shot noise has a Lorentzian line shape as can be seen in Fig.\,\ref{figure1}b.

To be able to extract the Fano factor we have done three different measurements. For each bias voltage, $V$, we have measured $\Gamma$ and the output noise from the amplifier, as we swept the gate charge, $Q_g$. This was done for 80 different bias voltages. To calibrate the amplifier noise we have in addition measured the amplifier output noise at zero bias as a function of gate charge.

The noise measurements were done at $f_{LC}$ with a bandwidth, $\Delta f=15\,\mathrm{kHz}$, significantly narrower then the resonance circuit bandwidth. In this case, the power spectral density of the shot noise can be expressed as a function of the magnitude of the reflection coefficient.
\begin{equation}{\label{Shot_noise}}
S_{shot}=\frac{g^2}{8}R\left(1-|\Gamma|^2\right)\langle i_s^2 \rangle=\frac{g^2}{4}\left(1-|\Gamma|^2\right)eFV,
\end{equation}
There is a clear advantage to express both the shot noise and the amplifier noise in terms of $[\Gamma |$ rather than the SET resistance, since parasitics of the tank circuit such as the shunt capacitance of the inductance is automatically included in $|\Gamma |$.

\begin{figure}[t]
\includegraphics[width=0.85\columnwidth]{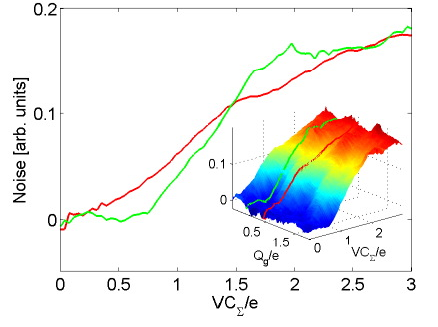}
\caption{\small (color online)  The shot noise generated by the single electron transistor as a function of bias voltage. The green curve shows the noise at maximum Coulomb blockade. The red curve shows the noise when the SET is in its open state. The amplifier noise has been subtracted as described in the text. The noise is measured at the resonance frequency in a bandwidth of  a $15\,$kHz. The inset shows a 3D-graph of the shot noise as a function of both bias voltage and gate charge.}
\label{Noise}
\end{figure}

\begin{figure}[t!]
\includegraphics[width=0.85\columnwidth]{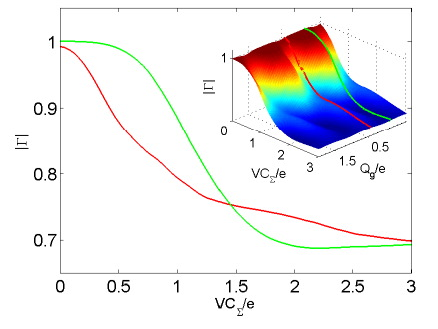}
\caption{\small (color online) The magnitude of the reflection coefficient as a function of the SET bias voltage. The green curve corresponds to the maximum of the Coulomb blockade. The red curve corresponds to the SET open state. The inset shows a 3D graph of $|\Gamma|$ as a function of both bias voltage and gate charge.}
\label{Gamma}
\end{figure}

The SET-generated noise for maximum Coulomb blockade, $Q_g=0$, and the SET open state, $Q_g=0.5 e$, is shown in Fig.\,\ref{Noise}. The inset shows the full set of data for all bias points. Here, the amplifier noise has been subtracted as will be explained later in the text. As can be seen there is no noise signal inside Coulomb blockade, when there is no current through the SET. For high bias voltages (data not shown), where the SET resistance reaches its asymptotic value, and $|\Gamma |$ remains constant, the noise increases linearly with bias.
As can be seen from eq.\,(\ref{Shot_noise}), we also need to measure $|\Gamma |$ at the same bias conditions, in order to extract the Fano factor.
Here we have normalized $|\Gamma |$ to unity in the Coulomb blockade region. In Fig.\,\ref{Gamma} we show the experimentally measured $|\Gamma |$ versus bias for maximum Coulomb blockade and for the SET open state. The insert shows the full bias - gate dependence. $|\Gamma|$ decreases monotonically from unity at low bias, to the asymptotic value $|\Gamma|=0.7$ at high bias. Thus, we conclude that the SET is operated in the over-coupled regime, where the impedance of the resonance circuit $Z=Z_{LC}^2/R$ is always smaller then $Z_0$.

The noise at the output of the amplifier also has a contribution from the amplifier itself, which should be accounted for. Generally the amplifier noise depends on the impedance of the resonance circuit. We describe the noise sources of the amplifier and refer them to the amplifier input \cite{Schiek}. In general two noise sources are required: a series voltage-noise source $v_n$, and a shunt current-noise source $i_n$, as shown in the Fig.\,\ref{figure1}a. The current noise source is due to the input current of the first HEMT transistor. The voltage noise source represents the internal amplifier noise referred to the input of the amplifier. The noise power at the output of the amplifier will clearly depend on the impedance attached to the input of the amplifier.
We can express also the amplifier noise as a function of $|\Gamma |$:
\begin{eqnarray}{\label{AmplNoise}}
S_{amp} =\frac{g^2}{8} [ \left( 1+|\Gamma|^2 \right) \left( \langle i_n^2\rangle Z_0 +\langle v_n^2\rangle/Z_0 \right)+ \nonumber \\
2|\Gamma | \cos(\phi-2\beta l) \left( \langle i_n^2\rangle Z_0 - \langle v_n^2\rangle/Z_0 \right)],
\end{eqnarray}
where $\phi$ is the phase of the voltage reflection from the resonance circuit, and $-2\beta l$ is the phase delay in the coaxial cable. Amplifiers are normally optimized so that the voltage noise and the current noise contributions are equally large for a source impedance of $Z_0$ and therefore the second term in eq.\,(\ref{AmplNoise}) is very small and can be neglected.
%
\begin{figure*}[t]
\begin{center}
\includegraphics[width= 1.8\columnwidth]{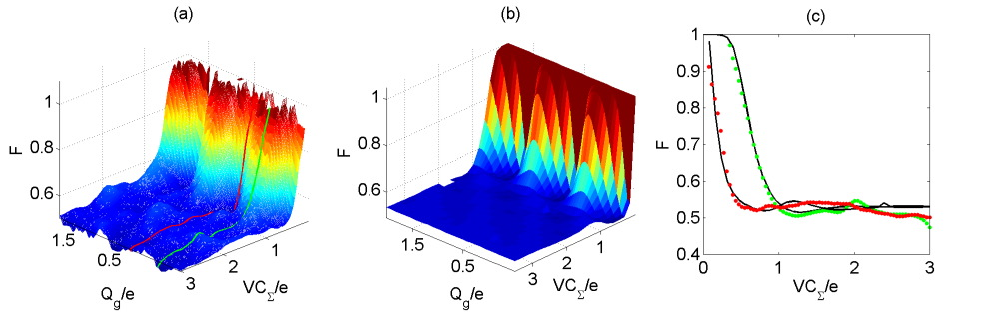}
\caption{\small (color online) The Fano factor (a) The measured Fano factor for the SET as a function of bias voltage $V$ and gate charge $Q_g$, the colored lines are the line-cuts shown in panel c for fixed $Q_g$. (b) The calculated Fano factor as a function of bias voltage $V$ and gate charge $Q_g$. The calculation is done using the orthodox theory \cite{PhysRevB.49.10381,Orthodox} for single electron tunneling, neglecting co-tunneling. We have assumed that transistor is overheated from the cryostat base temperature $T\approx 0.1E_c=250\,\mathrm{mK}$. (c) Measured Fano factor as a function of bias voltage for open state (red dots) and maximum Coulomb blockade (green dots). The black lines are the corresponding calculations using the orthodox theory \cite{PhysRevB.49.10381}.}
\label{Fano}
\end{center}
\end{figure*}
To calibrate the amplifier noise, the SET was biased at zero voltage, and the output amplifier noise was measured as a function of the SET gate charge, $\Gamma$ was measured for the same gate charges. Thus we can plot the amplifier noise as a function of $|\Gamma |$ in the range
($0.9<|\Gamma |<1$) as shown in Fig.\,\ref{figure1}c. The amplifier noise is well fitted by eq.\,(\ref{AmplNoise}) neglecting the second term.

The amplifier noise can then be subtracted from the measured noise so that we obtain the bare shot noise signal. To do this we extrapolate the data using eq.\,(\ref{AmplNoise}) so that we can subtract the noise in the whole range $0.7<|\Gamma |<1$.
Finally we combine eqs.\,(\ref{Shot_noise}) and (\ref{AmplNoise}) to extract the Fano factor over the hole range of bias and gate voltage, the result is plotted in Fig.\,\ref{Fano}a.
\begin{equation}{\label{Extract}}
F(V,Q_g)=\frac{S_{meas}(V,Q_g)-S_{amp}(|\Gamma |)}{\frac{g^2}{4}\left(1-|\Gamma|^2\right)eV}
\end{equation}
There is some uncertainty in the amplifier voltage gain $g$, and we have used the asymptotic value of the Fano factor at high bias to set the gain in Fig.\,\ref{Fano}a. The Fano factor asymptotic value is weakly dependent on the tunnel junctions resistances \cite{PhysRevB.49.10381}. Assuming that the resistance asymmetry is the same as the capacitances asymmetry in the tunnel junctions ($30\,\%$), the asymptotic Fano factor value should be: $F=(R_1^2+R_2^2)/(R_1+R_2)^2\approx 0.51$.

Fig.\,\ref{Fano}a represents the main result of our work, it shows a complete map of the Fano factor as a function of both bias voltage and gate charge. F is close to 0.5 at high bias and increases to unity as the bias is lowered towards the Coulomb threshold. Close to the threshold F varies between 0.5 and 1 as a function of gate charge. In the Coulomb blockade region both the shot noise and the current goes to zero and the Fano factor loses its meaning.

For comparison, we show in Fig.\,\ref{Fano}b the calculated Fano factor as function of bias voltages and gate charge, using the orthodox theory of single electron tunneling in the low frequency limit \cite{PhysRevB.49.10381,Orthodox}. In these calculations we have included 5 charge states on the SET island and we have neglected cotunneling. We have also taken into account that the SET is substantially overheated from the cryostat base temperature \cite{JApplPhys.78.2830}, using a SET temperature of $T\approx 0.1E_c=250\,\mathrm{mK}$ gives a very good agreement between experiment and theory.

Especially interesting cases are when the SET is fully open, $Q_g=0.5$, and when the Coulomb blockade is maximum, $Q_g=0$. These two cases are show in Fig.\,\ref{Fano}c, the red and green dots are experimental data for $Q_g=0.5$ and $Q_g=0$ respectively. The black lines are the calculations using the orthodox theory. 

The measurements shown here demonstrate the correlations of tunneling in the single electron transistor, which are well understood. At low voltages, the tunneling is limited by the rate in one of the junctions and the tunneling becomes correlated giving a Fano factor close to unity since the SET then behaves more like a single junction. At high bias the tunneling in the two junctions is sequential and uncorrelated, which results in the reduced shot noise $2e\langle I \rangle/N$. Using the method described here it should be possible to study also more complicated correlations in the SET including cotunneling and also combined Cooper-pair quasi-particle processes which occur when the SET is in the superconducting state.

In conclusion we have introduced a method for shot noise measurements of mesoscopic systems embedded in a resonance circuit. The measurements are done at a finite frequency where the $1/f$ noise can be neglected. We have been able to measure the shot noise generated by a Single Electron Transistor as a function of both bias and gate voltages. From the data we can extract the Fano factor which varies between $0.5$ and $1$ depending on the amount of correlation of the tunneling in the two junctions.  Our  experimental results agree very well with the orthodox theory for single electron tunneling.

We would like to acknowledge helpful discussions with J. Clarke, T. Duty, G. Johansson, and C. Wilson. The samples were made at the MC2 clean room at Chalmers. This work was supported by the Swedish SSF and VR, and by the Wallenberg foundation.

\bibliographystyle{apsrev}
\bibliography{References}
\end{document}